\documentclass[letterpaper, 10 pt, conference]{ieeeconf}

\IEEEoverridecommandlockouts                              

\usepackage[pdftex]{graphicx}
\usepackage{siunitx}
\usepackage{subfig}
\usepackage{cancel}
\usepackage{mathtools}
\usepackage{amsmath}
\usepackage{amssymb}
\usepackage{siunitx}
\newtheorem{lemma}{\textbf{Lemma}}
\newtheorem{remark}{\textbf{Remark}}

\title{\LARGE \bf
	Reset band for mitigatation of quantization induced performance degradation
}

\author{Bas Kieft, S. Hassan HosseinNia$^{*}$ and Niranjan Saikumar%
	\thanks{All authors are with Precision and Microsystems Engineering, 3ME, Delft University of Technology, The Netherlands}%
	\thanks{$^{*}$Corresponding author - \small{S.H.HosseinNiaKani@tudelft.nl}}%
	\thanks{The authors would like to thank Hittech Multin. This work was supported by NWO, through OTP TTW project $\#16335$.}
}

\begin{document}
	
\maketitle

\begin{abstract}
Reset control has emerged as a viable alternative to popular PID, capable of outperforming and overcoming the linear limitations. However, in motion control systems, quantization can cause severe performance degradation. This paper investigates this degradation in practical systems and re-purposes the reset band condition in order to mitigate the same. Numerical simulations have been conducted on a mass based positioning system to analyze the cause of the quantization induced performance degradation. Moreover, a performance and robustness analysis was performed with the proposed solution. Finally, novel tuning guidelines are provided for estimating the required reset band parameter. Additionally, practical experiments have been conducted on a high precision motion system for validation. The experiments show by example that the reset band can reduce the error in the problematic region by up to 285\% and hence shows the need and effectiveness of the proposed solution. 
\end{abstract}


\section{Introduction}
Reset control has gained popularity because it enables controller design that overcomes the limitations of linear control as posed by Bode. Reset control can be considered as a nonlinear alteration of PID control that reduces the coupling of gain and phase characteristics, thus diminishing Bode's limitation. Clegg first introduced reset with the Clegg Integrator (CI) \cite{Clegg2013}. This element is an integrator whose state is reset to zero when the input is zero. This reduces phase lag from $\SI{-90}{^\circ}$ to $\SI{-38}{^\circ}$. Later, as an extension of this idea, the First Order Reset Element (FORE) was created \cite{Horowitz1975}, with the Second Order Reset Element (SORE) \cite{Hazeleger2016} being the natural advancement of the FORE. In \cite{Hosseinnia2013}, a CI with fractional order is studied, thereby increasing the freedom of tuning further. 

A different take on tunable nonlinearity is applied with PI+CI and partial reset approaches. With PI+CI approach \cite{Ba2007}, a CI and a PI work in parallel with weighted gains provided to each branch. And with partial reset, the state is not reset to zero, but to a predetermined fraction of the state value at the time of reset. This lead to the Generalized FORE (GFORE) \cite{Guo2009} and the Generalized SORE (GSORE) \cite{Saikumar2019} in literature.

The benefits of reset control have been seen in process industry and motion control. In \cite{Vidal2010} a PI+CI element was applied to an industrial heat exchanger. Moreover, in this paper a variety of reset conditions are applied and investigated.
Hard Disk Drive (HDD), where positioning speed is crucial for performance is the focus of reset control research in \cite{Li2011a, Li2009a, Guo2009}. Similarly reset control has been applied for performance improvement in servo motors \cite{Hosseinnia2013}, tape-speed setup \cite{Zheng2000} and exhaust gas recirculation \cite{Panni2014}.

In the above stated applications, reset control has been used to decrease the phase lag introduced by an integrator. However, the decoupling of phase and gain allows for more interesting elements. The Constant in gain Lead in phase (CgLp) element is a prime example of the possibilities \cite{Saikumar2019,Palanikumar2018}. CgLp is designed by combining GFORE or GSORE with a corresponding order linear lead element and provides broadband phase lead without affecting the gain. This allows for a lowered dependency on the derivative component of PID.

One of the main advantages of reset control is that frequency domain loop-shaping for controller tuning is possible which is preferred by the industry. With the introduction of the CI, Clegg used a Sinusoidal Input Describing Function (SIDF) to linearize this reset element. This approach was adopted for all discussed elements \cite{Clegg2013,Horowitz1975,Hazeleger2016,Hosseinnia2013,Ba2007,Guo2009,Saikumar2019,Palanikumar2018}. While the use of SIDF has been improved upon to include harmonics in the form of Higher Order SIDF (HOSIDF) \cite{saikumar2020loop}, few works utilizing this description exist.

In almost all cases in literature, results extracted from simulations utilize continuous time implementations. In \cite{Palanikumar2018, Chen2019a, Chen2019, Saikumar, Saikumar2019}, the controllers were discretized. While practical results with discretization have been provided, in general, other practical issues have not been dealt with. Quantization is one such issue seen in practice that is not studied.

Quantization is a relevant signal disturbance that has been compared to noise. It has been researched extensively, ex. in \cite{Kaufmann2012,Linnenbrink2006,Lai2003}. Moreover, it's influence on linear control is well known \cite{Ferrante2015,Minyue2008,Park2019}. In linear control, quantization causes some mild limit cycling between quantization levels. Industry often prescribes the required resolution based on a safety factor times the desired precision. While sufficient works studying quantization in linear systems are present, research on the effect of quantization on reset control is lacking. Quantization with reset control is studied in \cite{Zheng2010}, but only for the application of reset control in the state estimator in order to mitigate quantization noise, known as distortion. For most reset control applications, the reset condition is a zero error crossing. Therefore, any distortion in the error can cause improper resets or missed resets. This could lead to performance degradation.

Although quantization is generally modeled as white or colored noise \cite{Zhu2013}, this does not represent reality. For example, the frequency of distortion is dependent on the quantizer resolution and response slope. If quantization in real-world systems and its effects are not considered, reset control can outperform PID as shown in \cite{Beker2001a}. However, quantization can cause performance degradation at frequencies for well below the bandwidth frequency, which is the region concerned with reference tracking and disturbance rejection. While performance in this region is affected, the main benefit of reset in terms of phase is desired in the region of the bandwidth. Therefore, the desire arises to have a linear controller at lower frequencies to mitigate/remove the quantization induced problems, and a reset controller at frequencies in the region of bandwidth and beyond.

This paper studies the use of reset bands to achieve this split. Instead of a reset at the zero error crossing, the reset band method places a band around the zero error \cite{Barreiro2011,Banos2012}. In this approach reset occurs when entering this band. The application of this reset condition is not widely studied besides the work done by the original authors. Additionally, no practical tuning rules exist for this reset condition.

The main contributions of this paper are a proposed solution for quantization induced performance degradation based on the application of a reset band in order to split the controller to perform linearly at lower frequencies and provide the reset action in the bandwidth region. Moreover, a method to tune the proposed solution and a describing function analysis with the inclusion of a reset band are provided.

The remainder of the paper is structured as follows: Section \ref{sec:preliminaries} states the preliminaries required for the paper. Then in section \ref{sec:QIPD}, the performance degradation is expanded upon. The reset band solution to counter the degraded performance is proposed in section \ref{sec:ResetBand}. The results are validated in section \ref{sec:Application} and the conclusions are given in section \ref{sec:Conclusion}.

\section{Preliminaries}\label{sec:preliminaries}

\subsection{Reset Control Definition}
In literature several definitions can be found referring to Reset Control. Zaccarian \cite{Janzamin2018} shows some of these definitions. For this research paper, the definition in line with \cite{Banos2012} will be followed. This definition has been expanded upon such that it leads to the form seen in (\ref{eq:resetdef}), e.g. \cite{Saikumar2019} and \cite{Guo2009}.
\begin{equation}\label{eq:resetdef}
R: \; \; 
\begin{cases} 
\dot{\mathbf{x}}_r(t) = A_r\mathbf{x}_r(t) + B_r e(t) & \text{if } e(t) \ne 0\\
\mathbf{x}_r(t^{+})=A_\rho\mathbf{x}_r(t) &  \text{if } e(t) = 0\\
u(t)=C_r\mathbf{x}_r(t)+D_r e(t)
\end{cases}
\end{equation}
where $\mathbf{x}_r(t)$ are the states of the reset controller, $u(t)$ is the output, e(t) is the error input, and $A_r$, $B_r$, $C_r$ and $D_r$ are the state-space matrices and are referred to as the base-linear system.  The first line of (\ref{eq:resetdef}) is referred to as the flow state, whereas  the second line is known as the jump state. $A_\rho$ is diagonal and is the reset matrix that determines the states to be reset and their after reset value. For a full reset, $A_\rho$ is the zero matrix and for a linear controller, this is the identity matrix.

\subsection{Describing Function}\label{sec:DF}
Reset control is non-linear. Therefore a Sinusoidal Input Describing Function (SIDF) analysis is to be performed such that the reset controller is approximated. In \cite{Guo2009} the describing function for the reset element of (\ref{eq:resetdef}) is presented as:
\begin{equation}\label{eq:G}
G(j\omega)=C^{T}_r(j\omega I-A_r)^{-1}(I+j\Theta_\rho(\omega))B_r+D_r
\end{equation}
where 
\begin{equation}\label{eq:Theta}
\Theta_\rho \overset{\Delta}{=} \frac{2}{\pi} \left( I+e^{\frac{\pi A_r}{\omega}}\right)\left(\frac{I-A_\rho}{I+A_\rho e^{\frac{\pi A_r}{\omega}}}\right)\left(\left(\frac{A_r}{\omega}\right)^{2}+I\right)^{-1}
\end{equation}

\subsection{Stability}
The $H_\beta$ condition from literature can be used to analyse the stability of a reset control system \cite{Beker2002}. A Reset control system (RCS) consisting of a reset controller defined in (\ref{eq:resetdef}) as its feedback controller for a plant with matrices $A_p, B_p, C_p \text{ and } D_p$ is quadratically stable, if there exists a $\beta \; \in \; \mathbb{R}^{n_r \times 1}$ and a positive definite $P_\rho \; \in \; \mathbb{R}^{n_r \times n_r}$ such that
\begin{align}
	\mathcal{H}_\beta \triangleq 
	\begin{bmatrix}
		\beta C_P & 0_{n_r \times n_{nr}}& P_\rho
	\end{bmatrix} \left(sI - A_{cl}\right)^{-1}
	\begin{bmatrix}
		0 \\ 0_{n_{nr} \times n_r}\\I_{n_r}
	\end{bmatrix}
\end{align}
is strictly positive real, where
$A_{cl}=\begin{bmatrix}
A_p & B_p C_r \\
-B_r C_p & A_r
\end{bmatrix}$\\
denotes the closed-loop $A$-matrix, $n_r$ indicates the number of states being reset, $n_{nr}$ indicates the number of non-resetting states. For partial reset, an additional condition given below needs to be satisfied.
$$A_\rho P_\rho A_\rho - P_\rho \leq 0$$

\subsection{Reset Elements and select controller designs}
\subsubsection{Clegg integrator}
Following the definition of (\ref{eq:resetdef}), the CI is defined with $A_r = 0$, $B_r = 1$, $C_r = 1$, $D_r = 0$ and $A_\rho = 0$. A generalization of CI results in $A_\rho \in [-1,1]$

\subsubsection{FORE}
Horowitz introduced a First Order Reset Element (FORE) \cite{Horowitz1975} to allow  for a reset first-order low-pass filter (LPF) with corner frequency $\omega_r$. Following the definition of (\ref{eq:resetdef}), the FORE is defined with $A_r = -\omega_r$, $B_r = \omega_r$, $C_r = 1$, $D_r = 0$ and $A_\rho = 0$. Generalized FORE (GFORE) has been defined in \cite{Guo2009} with $A_\rho = \gamma$ with $-1\le\gamma\le1$.

\subsubsection{SORE}
The Second Order Reset Element (SORE) was introduced in \cite{Hazeleger2016}. SORE has been defined according to (\ref{eq:resetdef}) with the following matrices:
\[A_r=\begin{bmatrix}
0 & 1 \\
-\omega_r^2 & -2\beta_r \omega_r
\end{bmatrix}, 
B_r = \begin{bmatrix}
0\\
\omega_r^2
\end{bmatrix},\]
\[C_r=\begin{bmatrix}
1 & 0
\end{bmatrix},
D_r=\begin{bmatrix}
0
\end{bmatrix}.\]
where $\beta_r$ is the damping coefficient. As with GFORE, SORE has also been generalized to include partial reset with a non-zero reset matrix $A_\rho = \gamma I$, where $-1\le\gamma\le1$ \cite{Saikumar2019}.

\subsubsection{Constant in gain Lead in phase element}
The reset element used in this paper is the first-order Constant in gain Lead in phase (CgLp) element \cite{Saikumar2019, Palanikumar2018}. This element is built by combining a GFORE element with a linear first-order lead. A second-order CgLp is possible by combining a SORE with a second-order linear lead.
\[
A_r=  \begin{bmatrix}
-\omega_{r\alpha} & 0\\
\omega_f & -\omega_f 
\end{bmatrix}, \quad
B_r=\begin{bmatrix}
\omega_{r\alpha}\\
0
\end{bmatrix},\]
\[
C_r=\begin{bmatrix}
\dfrac{\omega_f}{\omega_r} & \left(1-\dfrac{\omega_f}{\omega_r}\right)
\end{bmatrix},\quad
D_r=\begin{bmatrix}
0
\end{bmatrix},\quad
A_\rho = \begin{bmatrix}
\gamma & 0\\
0 & 1
\end{bmatrix}
\]
where $\omega_{r\alpha}$ , $\omega_f$ and $\omega_r$ are the corner frequency of FORE, starting and taming frequencies of the linear lead respectively. With CgLp, $\omega_{r\alpha}$ and $\omega_r$ are close to each other, resulting in a gain cancellation, but with phase lead in the \{$\omega_r,\omega_f$\} band.

\subsection{CgLp-PID}
The aforementioned CgLp element can be used in series with a linear PID controller to increase the phase margin resulting in the CgLp-PID controller. The design procedure is stated in \cite{Saikumar2019}. CgLp, having unity gain with phase lead, can be used to partially or completely replace the derivative part of PID and hence improve the overall loop-shape resulting in improved tracking, disturbance rejection and noise attenuation. A CgLp-PID controller can be represented as:
\begin{equation}\label{eq:CgLp-PID}
C= \underset{PI}{\underbrace{K(1+\frac{\omega_i}{s}})}\underset{D}{\underbrace{\frac{(s/\omega_d+1)}{(s/\omega_t+1)} }}\underset{CgLp}{\underbrace{\frac{1}{\cancelto{\gamma}{{s}/{\omega_{r\alpha}}+1}}\;\;\;\;\frac{{s}/{\omega_r +1}}{{s}/{\omega_f+1}} }}
\end{equation}
where the arrow indicates resetting action on the associated filter. The phase lead provided by linear derivative can be tuned by changing $a$ in $\omega_d = \omega_c/a$ and  $\omega_t = \omega_c a$. The phase lead provided by CgLp can be tuned by changing $\omega_r$ and $\gamma$.

\section{Quantization induced performance deterioration}\label{sec:QIPD}
\begin{figure}[h!]
	\centering
	\includegraphics[width=1\columnwidth]{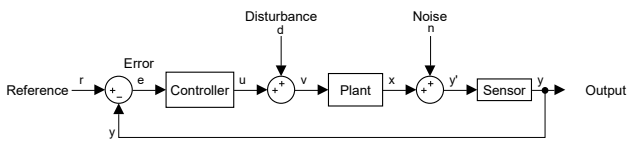}
	\caption{Block diagram of feedback control system with sensor introducing quantization.}\label{fig:BD}
\end{figure}

Consider the closed-loop control structure as shown in Fig. \ref{fig:BD}. Noise unrelated to quantization is added to the system output and results in $y'$, with quantization adding further noise.

\subsection{Distortion}
A good understanding of the error associated with quantization, known as distortion, can be gained from \cite{Linnenbrink2006}. Rounding quantization which is abundantly found in industrial products will be the focus of our work. A digital quantization level is defined as $Q=\frac{Range}{2^{bits}}$, where range is the maximum value that can be detected and bits is related to the digitization of the data. As an example, for a given sensor with a range of $\SI{1}{\mu m}$ a 5 bit and 6 bit systems will result in a resolution of $\frac{\SI{1000}{nm} }{2^5}= \SI{31.25}{nm}$ and $\frac{\SI{1000}{nm}}{2^6}=\SI{15.625}{nm}$ respectively. The boundary between two quantization levels in the sensed signal can be referred to as a jump or transition level. 

\subsection{Performance deterioration} \label{sec:RQLC}

In order to show performance deterioration due to quantization, a simulation study is performed. Consider a mass based positioning system with mass $m=\SI{1}{kg}$. A mass based system is chosen since the shape of the sensitivity function ($S$) is not affected by plant resonances or damping and deviations can be attributed to the controller. A CgLp-PID controller is tuned to get a bandwidth of $\SI{150}{Hz}$ with the controller parameters listed in Table \ref{tab:Settingsmass}. The considered quantization results in a resolution of $\SI{9.8}{\mu m}$. A linear approach to $S$ does not apply, because there is a nonlinear element in the loop resulting in harmonics in the output. Therefore the approach as defined in \cite{Cai} is adopted, see (\ref{eq:Ssigma}).
\begin{equation}\label{eq:Ssigma}
\begin{aligned}
&S_{\sigma}(\omega)=\frac{\max (|e(t)|)}{|R|}
&\text { for } t \geq t_{s s}
\end{aligned}\end{equation}

\begin{table}[]
	\centering
	\caption{Controller settings applied to the mass stage.}
	\label{tab:Settingsmass}
	{
			\begin{tabular}{|c|c|}
			\hline
			K                  & $\SI{6.0954e+05}{}$ \\ \hline
			$\omega_c$         & $\SI{942}{rad/s}$     \\ \hline
			$\omega_i$         & $\SI{94}{rad/s}$       \\ \hline
			$\omega_d$         & $\SI{530}{rad/s}$       \\ \hline
			$\omega_t$         & $\SI{1.68e+03}{rad/s}$  \\ \hline
			$\omega_{r\alpha}$ & $\SI{160}{rad/s}$      \\ \hline
			$\omega_f$         & $\SI{9.42e+03}{rad/s}$  \\ \hline
			$\omega_r$         & $\SI{172}{rad/s}$       \\ \hline
			$\gamma$           & $\SI{0.5}{}$     \\ \hline
			Range              & $\SI{5000}{\mu m}$         \\ \hline
			$F_s$			   & $\SI{10}{kHz}$        \\ \hline
		\end{tabular}%
	}
\end{table}

\begin{figure}
	\centering
	\includegraphics[trim = {1.5cm 0 2cm 0}, width=1\columnwidth]{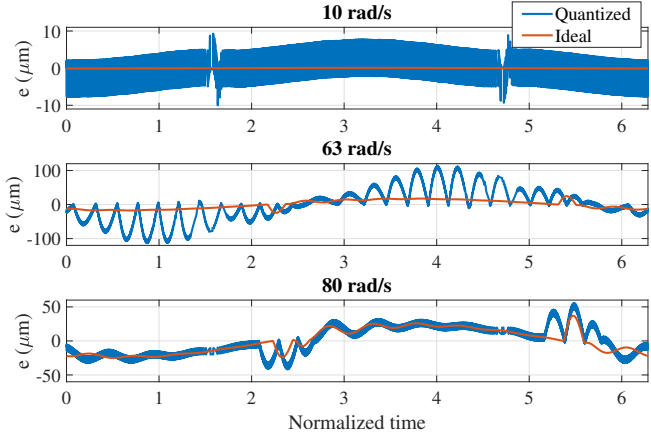}
	\caption{Steady state error response with and without quantization, where the sensor is quantized in the former case to provide a resolution of $\SI{9.8}{\mu m}$.}
	\label{fig:lastper}
\end{figure}

\begin{figure}
	\centering
	\includegraphics[width=1\columnwidth]{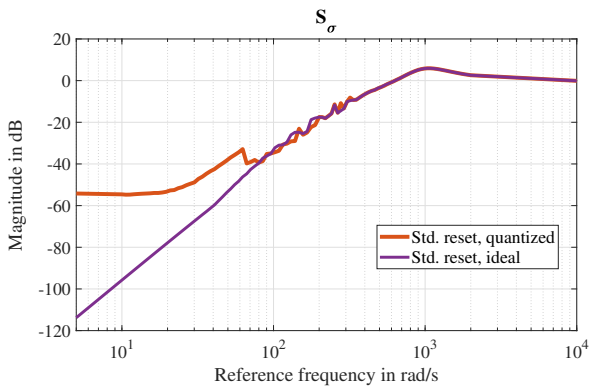}
	\caption{$S_\sigma$ for a mass system without and with quantization resulting in a resolution of $\SI{9.8}{\mu m}$.}\label{fig:Ssigmar0}
\end{figure}

Fig. \ref{fig:lastper} shows the steady state tracking error of the system for sinusoidal input at three frequencies in the absence and presence of quantization separately. Fig. \ref{fig:Ssigmar0} plots the $S_\sigma$ curve for the ideal and quantized cases. The first observation is related to the high $S_\sigma$ values at low frequencies in the quantized case which is not tending towards $-\infty$ dB as it is in the ideal case. Quantization creates a lower limit as seen in the time domain response in the first plot of Fig. \ref{fig:Ssigmar0} and $|S|$ will level out to $|\frac{1}{Q}|$. However, this is similar to the performance in linear systems in the presence of quantization. 

However, $S_{\sigma \; \text{quantized}}$ in Fig. \ref{fig:Ssigmar0} shows more deterioration. An increased magnitude with respect to $S_{\sigma \; \text{ideal}}$ is seen in the form of a bump around $\SI{63}{rad/s}$. The middle plot in Fig. \ref{fig:lastper} shows the response in the temporal domain. An analysis of the data shows that the increased magnitude is because of quantization induced excessive resetting.

In the presence of quantization, the zero crossing error reset condition provided in (\ref{eq:resetdef}) is satisfied when the mean of error is within $\frac{1}{2} Q \ge e \ge -\frac{1}{2} Q$. Within this error band, quantization will cause zero-error crossings which are not present in the ideal case and these excessive unmodelled resets lead to performance degradation.

\section{Reset band - solution for quantization induced performance deterioration}\label{sec:ResetBand}
\begin{figure}
	\centering
	\includegraphics[width=1\columnwidth]{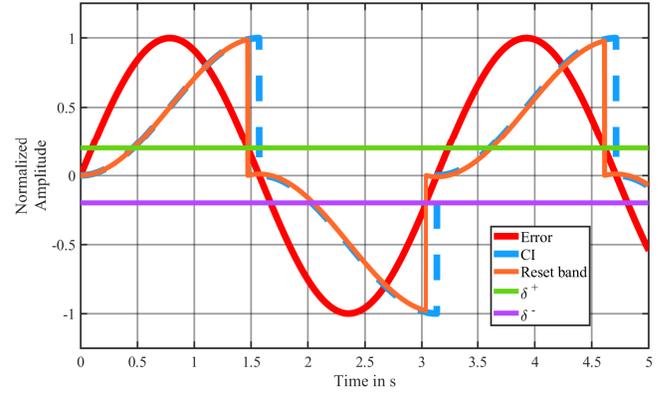}
	\caption{The response of a Clegg integrator with and without reset band.}\label{fig:resetband}
\end{figure}

The reset band condition is a special case which modifies the reset surface. Reset only occurs when the error enters the reset band. Fig. \ref{fig:resetband} shows an example where a reset band is applied to a CI. The standard CI resets when its input, the error, crosses zero. The CI with the reset band is seen to reset when the error enters the area of $|\delta|$ around zero.

\subsection{Reset band preliminaries}

\subsubsection{Definition}
The reset band controller is defined as in (\ref{eq:RBdef}) in \cite{Banos2012}.
\begin{equation}\label{eq:RBdef}
R_\delta \; \; 
\begin{cases} 
\dot{\mathbf{x}}_r(t) = A_r\mathbf{x}_r(t) + B_r e(t) & {\text { if }(e(t), \dot{e}(t)) \notin B_{\delta}}\\
\mathbf{x}_r(t^{+})=A_\rho\mathbf{x}_r(t) &  {\text { if }(e(t), \dot{e}(t)) \in B_{\delta}}\\
u(t)=C_r\mathbf{x}_r(t)+D_r e(t)
\end{cases}
\end{equation}
where
$$
\mathrm{B}_\delta=\{(x, y) \in \mathrm{R}^2 |(\mathrm{x}=-\delta \wedge \mathrm{y}>0) \vee(\mathrm{x}=\delta \wedge \mathrm{y}<0).
$$

Reset band elements will be denoted with the subscript $\delta$. So, a reset band FORE will be FORE$_\delta$, and FORE$_0$ represents the standard reset condition (with no reset band) as in FORE.

\subsubsection{Sinusoidal Input Describing Function (SIDF)}\label{sec:RBDF}
The reset band system can also be linearised with a SIDF. There is one major difference. For a reset band system, the SIDF is a function of both input frequency and amplitude. The SIDF for a full reset FORE$_\delta$ with input $e = E\sin{(\omega t)}$ is given in \cite{Banos2012}. However, in order to generalize the DF such that partial reset is included, the method of \cite{Cai} has been applied in this paper to provide the equations to analytically obtain the SIDF for any reset controller of form in (\ref{eq:RBdef}). 
\begin{equation} \label{eq:GeneralizedDF}
G(j\omega)=C_{r}\left(j \omega I-A_{r}\right)^{-1}\left(I+e^{j \phi} j \Theta_{s}(\omega)\right) B_{r}+D_{r}
\end{equation}
where
$\Theta_{s}=\Theta_{\rho}\left(\dfrac{-A_{r} \sin \phi+\omega \cos \phi I}{\omega}\right)$
and $\Theta_\rho$ is provided in Eq \ref{eq:Theta} with $\phi = \pi - \sin ^{-1}{\left(\dfrac{\delta}{E}\right)}$.

\subsubsection{Stability}
A stability analysis of reset control systems with a reset band is conducted in \cite{Banos2009}. An important note is that when $\dfrac{\delta}{E}$ tends towards 1, limit cycling might occur. 

\subsection{Performance analysis with reset band}
The same mass system and controller are considered to analyse performance with the reset band. A sinusoidal reference with an amplitude of $\SI{5000}{\mu m}$ and frequency of $\SI{50}{rad/s}$ was applied. In the absence of quantization, the error is noted to be $\SI{10}{\mu m}$. Hence $\delta$ is designed such that it is bigger than the maximum error in addition to $Q$. This leads to a required reset band of $\delta \approx \SI{20}{\mu m}$, which was applied. An additional simulation was conducted with a wrong value of $\delta \approx \SI{10}{\mu m}$.

\begin{figure}
	\centering
	\includegraphics[width=1\columnwidth]{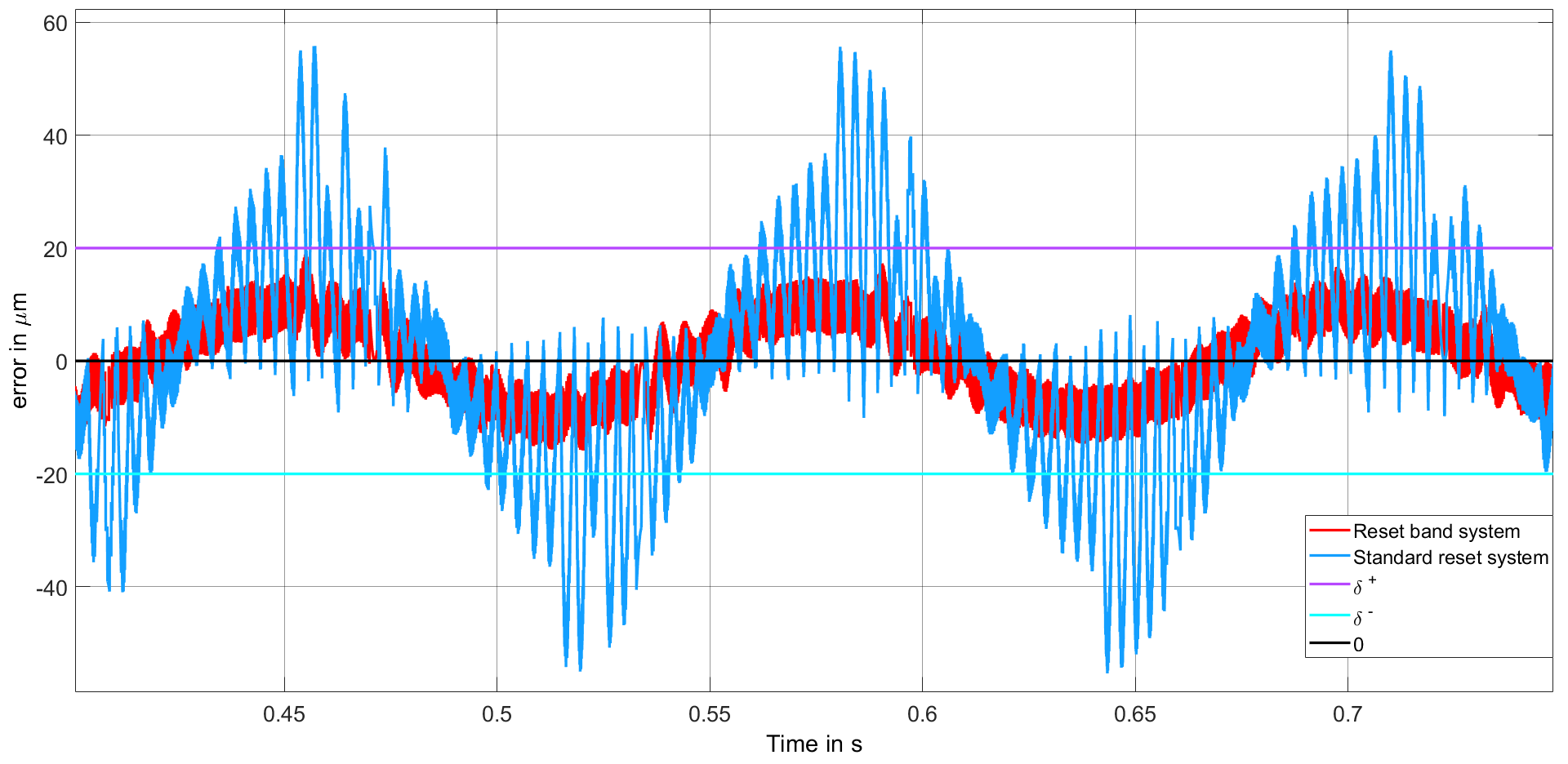}
	\caption{Steady state response of a mass system with and without reset band with the reset band tuned correctly.}\label{fig:rb_improvement}
\end{figure}

\begin{figure}
	\centering
	\includegraphics[width=1\columnwidth]{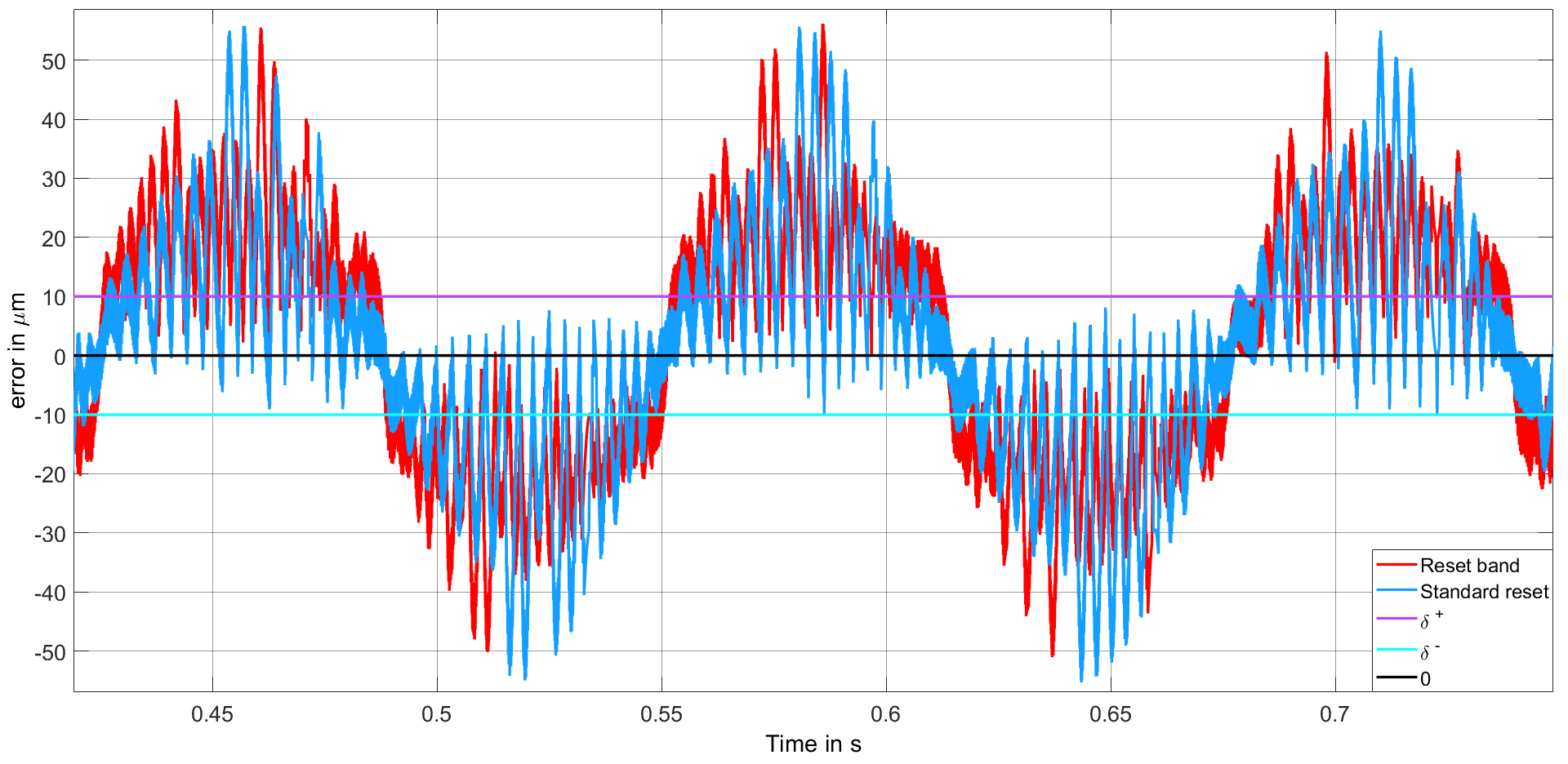}
	\caption{Steady state response of a mass system with and without reset band with the reset band tuned incorrectly.}\label{fig:rb_deter}
\end{figure}

While Fig. \ref{fig:rb_improvement} shows significant performance improvement with correctly chosen reset band, the time response in red in Fig. \ref{fig:rb_deter} shows that the reset band solution can cause performance deterioration if improperly tuned. The wrong tuning of $\delta$ can occur because of any of the following reasons. (i) $\delta$ was originally tuned for a lower frequency reference signal. With increase in frequency, there is a corresponding increase in the error even in the ideal case. (ii) $\delta$ is tuned for the wrong reference amplitude. If the reference amplitude increases, error increases. This increased error is able to cross the reset band and therefore resets will be induced and further due to quantization, the initial problem of unnecessary resets and performance deterioration is reintroduced. This is as seen in Fig. \ref{fig:rb_deter}.

The next point of consideration is that for higher frequencies close to the bandwidth, the ratio of $\frac{E}{\delta}$ will tend towards 1. According to \cite{Banos2009} this can cause limit cycling. Therefore proper tuning is essential and hence tuning guidelines have been provided in the next section.

\subsection{Tuning guidelines}\label{sec:ROT}
From Fig. \ref{fig:Ssigmar0} , we can see that quantization induced resets mainly create problems at frequencies significantly lower than the bandwidth. At these frequencies, the magnitude of error is comparable to the quantization levels. This leads to the desire to have a linear controller for lower frequencies and a reset controller around bandwith to get the phase advantage, which can be achieved by appropriate tuning of the reset band.

From theory the maximum quantization can be derived \cite{Linnenbrink2006}.
\begin{lemma}\label{lem:rQ}
	\begin{itemize}
		\item Rounding quantization can cause a maximum error of $\pm\frac{1}{2}Q$ relative to the true error.
 		\item Truncation quantization can cause a maximum error of $+Q$ relative to the true error.	
 		\end{itemize}
\end{lemma}

With Lemma \ref{lem:rQ} a linear range in the frequency domain can be identified and $\delta$ tuned to ensure this is based on a reference amplitude, as provided in Lemma \ref{lem:w<ws}.
\begin{lemma}\label{lem:w<ws}
	There will be no reset for $\omega < \omega_s$ if the following condition is satisfied:
	\begin{equation}\label{eq:bdelta_S}
	\delta=\text{max}{\left(e\right)|}_{\omega_s}+\frac{Q}{2}
	\end{equation}
	where: $\text{max}{\left(e\right)|}_{\omega <\omega_s}=|S|_{\text{BLS}}\cdot|r|$ and $|S|_{BLS}$ is the sensitivity function in the case the reset action is completely disabled.
\end{lemma}
\begin{remark}\label{rem:rtQ}
	Since a controller only responds to the distorted signal $y$, which is identical for both quantization types, Lemma \ref{lem:w<ws} holds both for truncation and rounding quantization. 
\end{remark}

\subsubsection{Noise and disturbances}
Noise is a common presence in practice. If $\delta$ is tuned too tightly, noise can induce error crossings through $\delta$. To resolve this a safety factor can be implemented, or a noise identification can be used to find the maximum noise amplitude which can then be added as the safety factor to $\delta$. Disturbances can be a problem as well. Disturbances can in most cases be measured beforehand to obtain an estimate of the maximum errors created, but it can also be deduced if a disturbance model is available.

The process sensitivity function $\frac{e}{d}=PS$ can be used in order to estimate the error as a result of the disturbance. Plant uncertainties along with the fact that the DF is used to predict the error means that a safety margin should always be applied to $\delta$:
\[\delta_{robust}=k\delta.\]
Where $k$ is a safety factor of $k \ge 1$. From here on, $\delta$ is the applied reset band magnitude with the safety margin considered.

\subsubsection{Multiple Sinusoid Reference / Sinusoidal Decomposition of Reference}
In practice there are many situations where a reference is composed of multiple sines, e.g. a 4th order trajectory \cite{Lambrechts}. Therefore it is imperative that this solution holds for superposition. Because a reset system is nonlinear, superposition does not hold. However, with the designed tunable linearity range, the following remark can be made:
\begin{remark}
	Within the reset band, the system is linear. Hence, as long as the error is within the reset band, superposition holds. 
\end{remark}

\begin{table}[]
	\centering
	\caption{Controller settings applied to the mass based positioning system}
	\label{tab:Settings2b}
	{
		\begin{tabular}{|l|l|l|}
			\hline
			P                  & 6.0954e+05  & -     \\ \hline
			$\omega_c$         & 942      & rad/s    \\ \hline
			$\omega_i$         & 94       & rad/s \\ \hline
			$\omega_d$         & 530      & rad/s \\ \hline
			$\omega_t$         & 1.68e+03 & rad/s \\ \hline
			$\omega_{r\alpha}$ & 160      & rad/s \\ \hline
			$\omega_f$         & 9.42e+03 & rad/s \\ \hline
			$\omega_r$         & 172      & rad/s \\ \hline
			$\gamma$           & 0.5      & -     \\ \hline
			Range              & 5000     & $\mu$ m    \\ \hline
			f$_\text{sampling}$& 10     & KHz    \\ \hline
			T$_{simulink}$	   & 10/f$_\text{sampling}$ & s \\ \hline 
		\end{tabular}%
	}
\end{table}

\begin{figure}
	\centering
	\includegraphics[width=1\columnwidth]{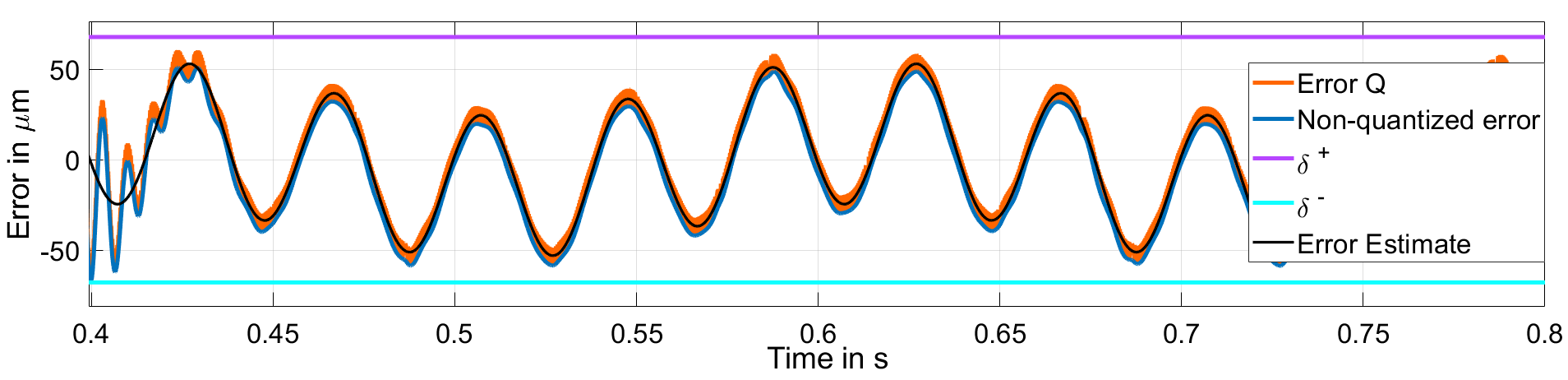}
	\caption{Error for a double sine reference.}\label{fig:doublesine}
\end{figure}

Figure \ref{fig:doublesine} shows the steady state response for the controller of Table \ref{tab:Settings2b} applied to the mass system. It can be seen in Fig. \ref{fig:doublesine} that the ideal non-quantized error, shown in blue, has $-\frac{1}{2} Q$ offset w.r.t. the estimated error, shown in black. This is due to truncation quantization. The reference was a summation of two sines with $A_1=5000 \mu$m, $A_2=\frac{1}{3} \cdot 5000 \mu$m, $f_1=5$ Hz and $f_2=25$ Hz. The sensitivity function was used to estimate the error for each individual sine which were summed. Hence, a Fourier decomposition can be used to convert any tracking reference to sines and this method employed for determining the correct value of $\delta$.

\subsection{Numerical sensitivity function}
\begin{figure}
	\centering
	\includegraphics[width=1\columnwidth]{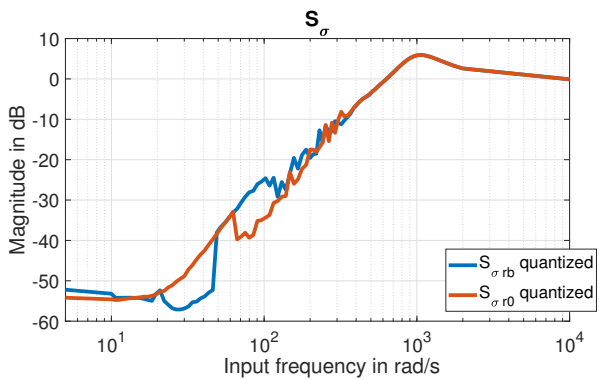}
	\caption{$S_\sigma$ with quantization for reset band and standard reset with reset band = 0.}\label{fig:Sqr0rb}
\end{figure}

The same system is also used to obtain the sensitivity function with the new reset band formulation. Fig. \ref{fig:Sqr0rb} shows the performance of the solution. The $\delta$ is tuned using the procedure described in Lemma \ref{lem:w<ws} for $\omega_s = \SI{50}{rad/s}$. It can be seen that an improvement is achieved at lower frequencies. Moreover, the end of the linear frequency range can be clearly seen in the large increase in $S_\sigma$ at $\omega_s\approx \SI{50}{rad/s}$.

\subsection{Sensitivity to $\delta$}
\begin{figure}
	\centering
	\includegraphics[width=1\columnwidth]{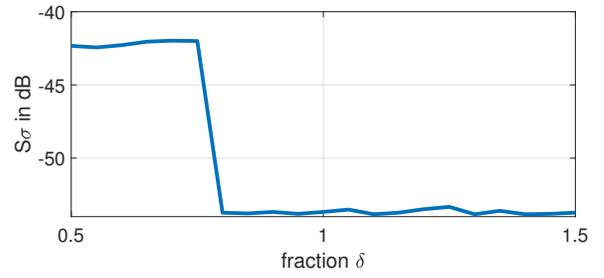}
	\caption{$S_\sigma$ vs delta for a quantized reference.}\label{fig:Sqdelta}
\end{figure}

To analyze the sensitivity of the tuning method, a series of simulations was performed. A sinusoid reference of with $A=\SI{5000}{\mu m}$ and $f \approx \SI{6.4}{Hz}$ was applied. $\delta$ was calculated according to (\ref{eq:bdelta_S}) to be $\SI{16}{\mu m} \approx \SI{5000}{\mu m} \cdot 0.00117 +\dfrac{1}{2}\cdot\dfrac{\SI{5000}{\mu m}}{2^9}$, where $0.00117$ is the value of $|S|_{BLS}$ at that frequency. The results are plotted in Fig. \ref{fig:Sqdelta} where it can be seen that there is a hard border where $\delta$ is suddenly no longer beneficial. This is the case when the error starts to hit the reset band at steady state. This hard edge is also visible in Fig. \ref{fig:Sqr0rb}. This shows that the solution of reset band  is not robust if the reference frequency is close to $\omega_s$.

\section{Practical validation}\label{sec:Application}

\subsection{Precision positioning stage}
\begin{figure}
	\centering
	\includegraphics[width=1\columnwidth]{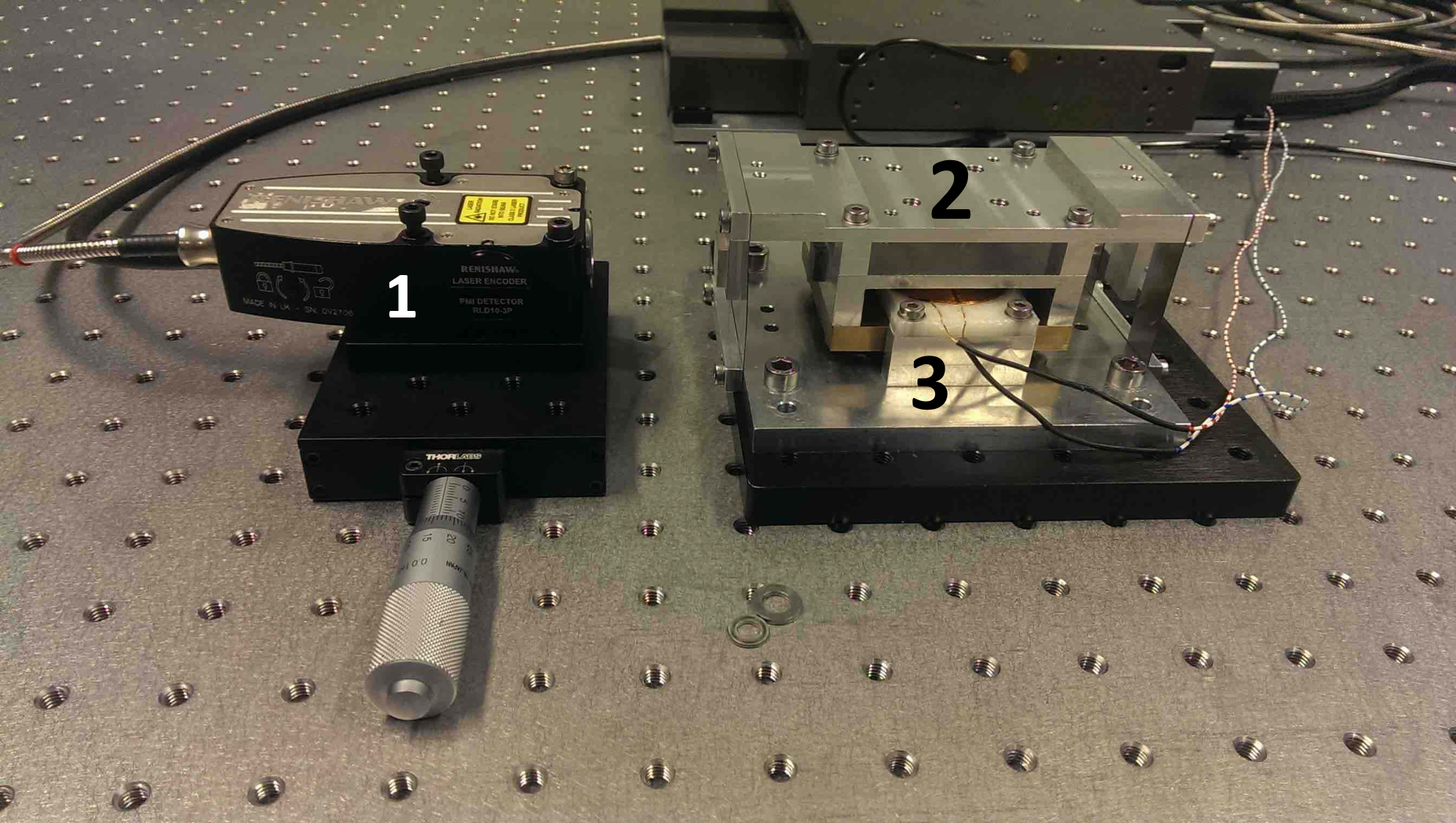}
	\caption{Fine positioning stage. 1: sensor, 2: mass stage, 3: actuator.}\label{fig:finestage}
\end{figure}

The system used for validation is a custom designed one-degree-of-freedom high precision positioning stage, see Fig. \ref{fig:finestage}. Essentially it is a mass-spring-damper system. It has a Lorentz actuator and a Renishaw RLE10 laser encoder set to $\SI{10}{nm}$ resolution. An FPGA NI cRIO system was utilized to achieve fast real-time control with a sampling rate of \SI{10}{kHz}.

\begin{figure}
	\centering
	\includegraphics[width=1\columnwidth]{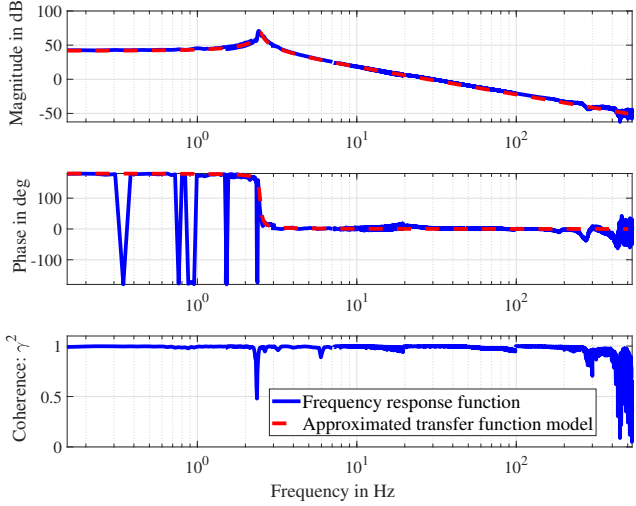}
	\caption{Frequency response data and estimated model.}\label{fig:Tfestimate}
\end{figure}

As is common in industry, frequency response data of the system is obtained by applying chirp signals. The results can be seen in Fig. \ref{fig:Tfestimate}. The following transfer function was estimated:
$$P(s)=\frac{3.038e04}{s^2 + 0.7413 s + 243.3}$$

\subsection{Designed controller}

\begin{table}[]
		\centering
	\caption{Controller settings for practical application.}
	\label{tab:SettingsCgLp}
		\begin{tabular}{|c|c|}
			\hline
			K                  & \SI{16.41}{}     \\ \hline
			$\gamma$           & \SI{0}{}         \\ \hline
			$\omega_c$         & \SI{942.5}{rad/s} \\ \hline
			$\omega_i$         & \SI{94.25}{rad/s} \\ \hline
			$\omega_d$         & \SI{529.2}{rad/s} \\ \hline
			$\omega_t$         & \SI{1679}{rad/s}  \\ \hline
			$\omega_{r\alpha}$ & \SI{697.6}{rad/s} \\ \hline
			$\omega_r$         & \SI{812.1}{rad/s} \\ \hline
			$\omega_f$         & \SI{9420}{rad/s}  \\ \hline
		\end{tabular}
	\end{table}

A CgLp-PID controller for a bandwidth (defined as crossover frequency) of \SI{150}{Hz} and phase margin of $40^{\circ}$ is designed using the transfer function of (\ref{eq:CgLp-PID}) with the parameters shown in Table \ref{tab:SettingsCgLp}. The controller was discretized according to the tustin method. 

\subsection{Results}
To show that the quantization induced performance degradation does not occur for all frequencies for all values of $Q$, two $Q$'s have been implemented. The higher resolution corresponds to \SI{10}{nm}. A second resolution was chosen to be \SI{80}{nm}.

\subsubsection{\SI{10}{nm} resolution}
\begin{figure}
	\centering
	\includegraphics[width=1\columnwidth]{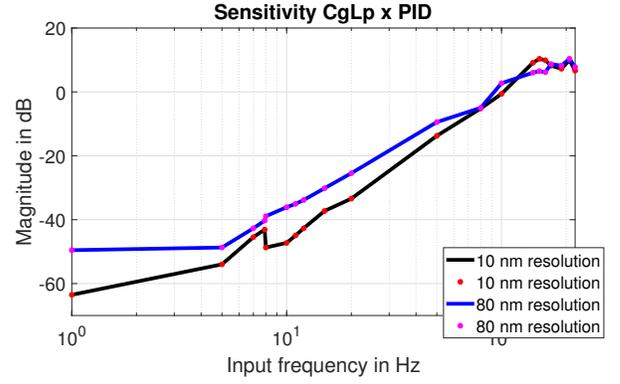}
	\caption{Experimentally deduced sensitivity functions for CgLp-PID with \SI{10}{nm} and \SI{80}{nm} resolutions.}\label{fig:S_CgLp_exp_Q0_Q3}
\end{figure}

For the low frequencies, a reference amplitude of $\SI{30}{\mu m}$ was applied. This leads to a theoretical limit of $\SI{-69}{dB}$ for a steady state \SI{10}{nm} error. In practice due to the presence of noise and disturbance in the form of floor vibrations, the limit was considered to be at \SI{20}{nm} at \SI{-63}{dB}. As expected, the quantization induced degradation occurs at the low frequency range. Fig \ref{fig:S_CgLp_exp_Q0_Q3} shows the sensitivity $S_\sigma$, and a bump similar to Fig. \ref{fig:Sqr0rb} can be seen to start from \SI{5}{Hz} and drop at \SI{8}{Hz} where the problematic region ends.

\begin{figure}
	\centering
	\includegraphics[width=1\columnwidth]{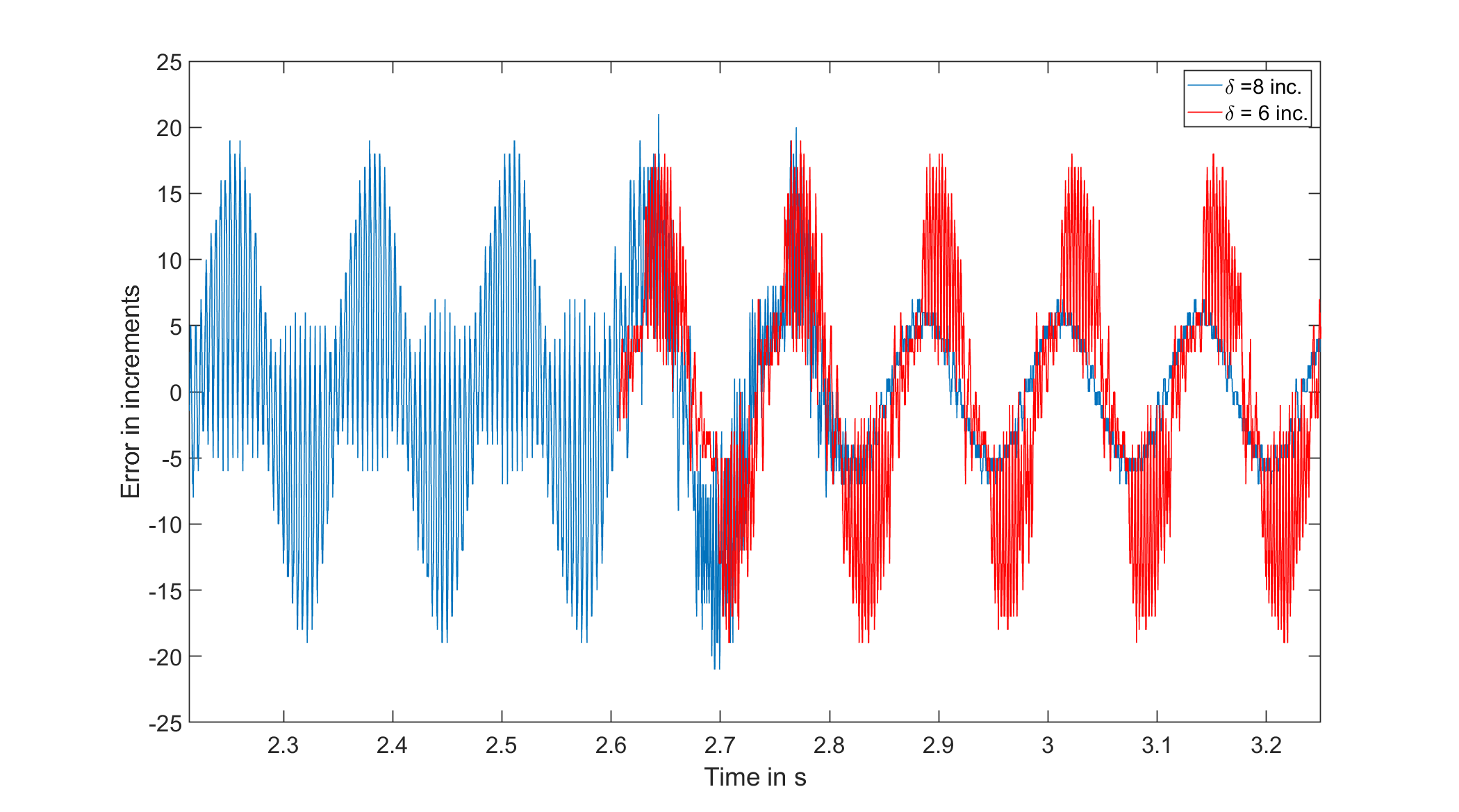}
	\caption{Steady state time response at 7.9 Hz, 1 inc. resolution which corresponds to 10nm. At $t=\SI{2.6}{s}$ reset band is initiated.}\label{fig:r0_good_bad_f7_9}
\end{figure}

In Fig. \ref{fig:r0_good_bad_f7_9}, the steady state time response for a sinusoidal reference of \SI{7.9}{Hz} is shown in blue. It can be seen that at first the error is \SI{200}{nm} (\SI{-43.5}{dB}) due to excessive resetting. At \SI{2.75}{s} a reset band of $\delta=\SI{80}{nm}$ is initiated, and hence the error reduces to \SI{70}{nm} (\SI{-52.6}{db}). Therefore for this frequency the reset band solution against quantization induced performance degradation leads to an improvement of \SI{9}{dB}. As can be seen in red in Fig. \ref{fig:r0_good_bad_f7_9}, the response with a mistuned $\delta = \SI{60}{nm}$ shows performance degradation. The response is only plotted from when the reset band was initiated. In this case no benefits from the reset band are found. However neither are the drawbacks as detrimental as seen in the earlier simulations.

\subsubsection{\SI{80}{nm} resolution}
\begin{figure}
	\centering
	\includegraphics[width=1\columnwidth]{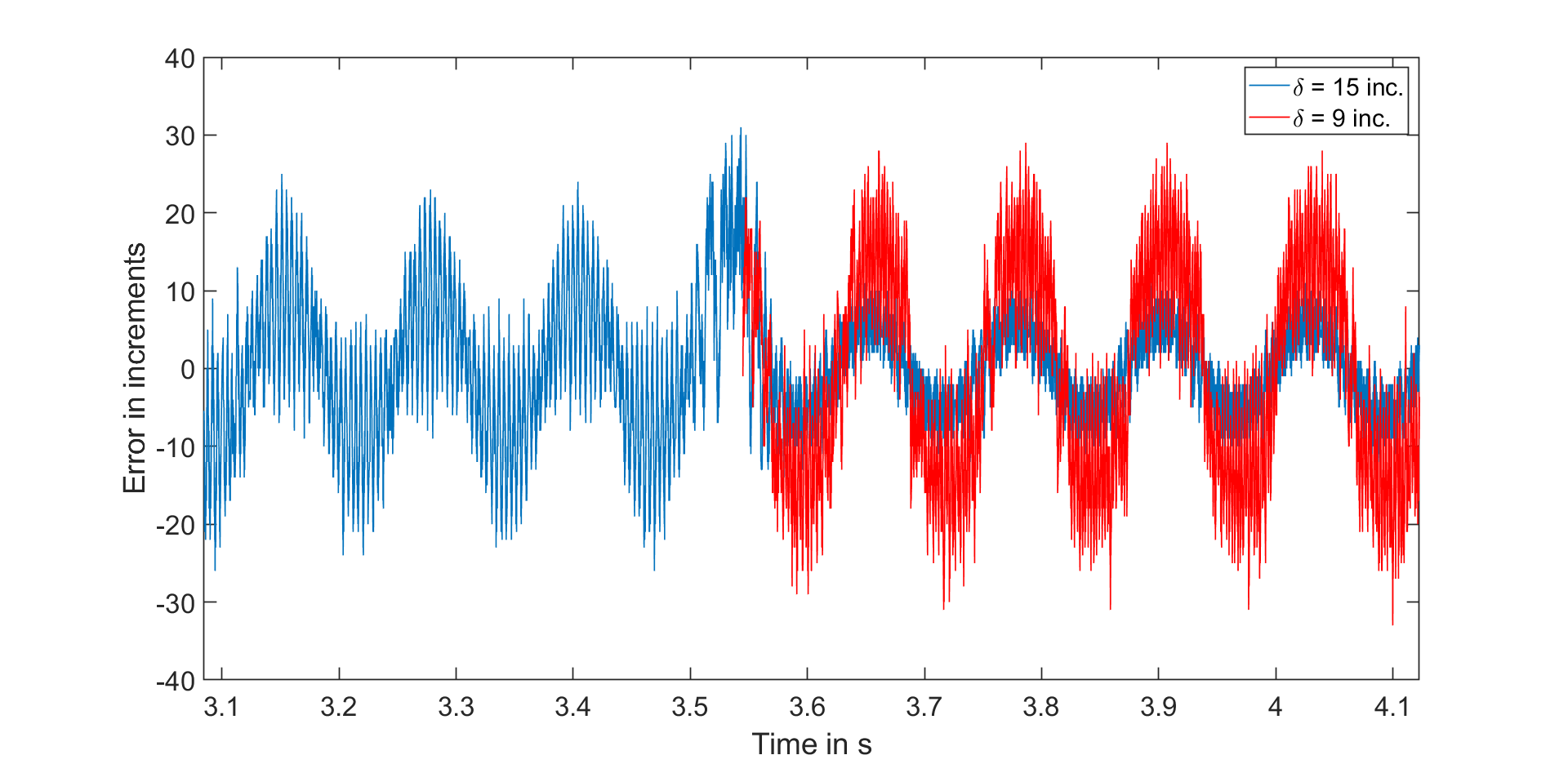}
	\caption{Steady state time response at 7.9 Hz, 8 inc. resolution which corresponds to 80nm. At $t=\SI{3.5}{s}$ reset band is initiated. }\label{fig:Q3_r0_good_bad_f7_9}
\end{figure}

Once again a reference amplitude of $\SI{30}{\mu m}$ was applied for lower frequencies. The theoretical limit is then \SI{-51.5}{dB}. In this case the theoretical and practical limit nicely agree because the noise level is small relative to the \SI{80}{nm} resolution. In Fig. \ref{fig:S_CgLp_exp_Q0_Q3}, the deduced sensitivity function for a standard reset condition is shown. In this case no bump can be seen, however quantization still induced performance degradation as shown in Fig. \ref{fig:Q3_r0_good_bad_f7_9}. Implementing a reset band with $\delta = \SI{150}{nm}$ results in an improvement of \SI{6}{dB}. As can be seen in the same figure, for this resolution a badly tuned reset band will cause performance degradation. A degradation of \SI{2}{dB} was recorded. Once again the response of the second $\delta$ was only plotted after initiation.

\section{Conclusion}\label{sec:Conclusion}
Reset control is popularly used and has been studied in literature on several motion and process control applications. However, there has been no focus on the practical issue of sensor signal quantization and its effect on performance. In this paper it is shown that quantization can degrade performance of reset controllers.

Reset band is proposed as a solution for quantization induced performance degradation. Tuning guidelines are provided. Experiments have validated that a reset band can greatly reduce quantization induced performance degradation. In one case the error was reduced from \SI{200}{nm} to less than \SI{80}{nm}. However, when improperly tuned the reset band can cause further performance degradation.

An interesting direction of research would be to investigate whether real time adapted $\delta$, e.g. in the form of a variable or advanced reset band, could improve quantization performance degradation farther.  Also, the implementation of an observer can reduce distortion and thus result in a smaller value for $\delta$.

\bibliographystyle{IEEEtran}
\bibliography{library}

\end{document}